\documentclass[11pt]{article}
 \usepackage{geometry}                
 \usepackage{mathrsfs,graphicx,amsmath,amssymb,amsbsy}
 \title{Asymptotic Freedom: From Paradox to Paradigm\footnote{Lecture given in acceptance of the Nobel Prize, Dec. 2004}}
 \author{Frank Wilczek}
 \date{\small\it MIT-CTP-3605}                                           
 
\begin{document}
 \maketitle

 \section{A Pair of Paradoxes}
 
In theoretical physics, paradoxes are good.  That's paradoxical,  since a paradox  appears to be a contradiction, and contradictions imply serious error.   
But Nature cannot realize contradictions.  When our physical theories lead to paradox we must find a way out.   Paradoxes focus our attention, and we think harder.

When David Gross and I began the work that led to this Nobel Prize \cite{disc, gw1, gw2, gw3}, in 1972, we were driven by paradoxes.  In resolving the paradoxes we were led to discover a new dynamical principle, asymptotic freedom.  This principle in turn has led to an expanded conception of fundamental particles, a new understanding of how matter gets its mass, a new and much clearer picture of the early universe, and new ideas about the unity of Nature's forces.   Today I'd like to share with you the story of these ideas.

 \subsection{Paradox 1: Quarks are Born Free, but Everywhere They are in Chains}
 
The first paradox was phenomenological.   

Near the beginning of the twentieth century, after pioneering experiments by Rutherford, Geiger and Marsden, physicists discovered that most of the mass and all of the positive charge inside an atom is concentrated in a tiny central nucleus.   In 1932 Chadwick discovered neutrons, which together with protons could be considered as the  ingredients out of which atomic nuclei could be constructed.   But the known forces, gravity and electromagnetism, were insufficient to bind protons and neutrons tightly together into objects as small as the observed nuclei.   Physicists were confronted with a new force, the most powerful in Nature. It became a major challenge in fundamental physics, to understand this new force.

For many years physicists gathered data to address that challenge, basically by bashing protons and neutrons  together and studying what came out.  The results that emerged from these studies, however, were complicated and hard to interpret.

What you would expect, if the particles were really fundamental (indestructible), would be the
same particles you started with, coming out with just their trajectories changed.
Instead,  the outcome of the collisions was often many particles.  
The final state might contain several copies of the originals, or different particles altogether.  A plethora of new particles was discovered in this way.  Although these particles, generically called hadrons, are unstable, they otherwise behave in ways that broadly resemble the way protons and neutrons behave.  So the character of the subject changed.  It was no longer natural to think of it as simply as the study of a new force that binds protons and neutrons into atomic nuclei.  Rather, a new world of phenomena had come into view.  This world contained many unexpected new particles, that could transform into one another in a bewildering variety of ways.   Reflecting this change in perspective, there was a change in terminology.   Instead of the nuclear force, physicists came to speak of the strong interaction.

In the early 1960s, Murray Gell-Mann and George Zweig made a great advance in the theory of the strong interaction, by proposing the concept of quarks.  If you imagined that hadrons were not fundamental particles, but rather that they were assembled from a few more basic types, the quarks, patterns clicked into place.  The dozens of observed hadrons could be understood, at least roughly, as 
different ways of putting together just three kinds (``flavors'') of quarks.  You can have a given set of quarks in different spatial orbits, or with their spins aligned in different ways.  The energy of the configuration will depend on these things, and so there will be a number of states with different energies, giving rise to particles with different masses, according to $m=E/c^2$.  It is analogous to the way we understand the spectrum of excited states of an atom, as arising from different orbits and spin alignments of electrons.   
(For electrons in atoms the interaction energies are relatively small, however, and the effect of these energies on the overall {\it mass\/} of the atoms is insignificant.)

The rules for using quarks to model reality seemed quite weird, however. 

Quarks were supposed to hardly notice one another when they were close together, but if you tried to isolate one, you found that you could not.  People looked very hard for individual quarks, but without success.   Only bound states of a quark and an antiquark -- mesons -- or bound states of three quarks -- baryons -- are observed.  This experimental regularity was elevated into The Principle of Confinement.  But giving it a dignified name didn't make it less weird.

There were other peculiar things about quarks.  They were supposed to have electric charges whose magnitudes are fractions ($\frac{2}{3}$ or $\frac{1}{3}$) of what appears to be the basic unit, namely the magnitude of charge carried by an electron or proton.  All other observed electric charges are known, with great accuracy, to be whole-number multiples of this unit.   Also, identical quarks did not appear to obey the normal rules of quantum statistics.  These rules would require that, as spin $\frac{1}{2}$ particles, quarks should be fermions, with antisymmetric wave functions.   The pattern of observed baryons cannot be understood using antisymmetric wave functions; it requires symmetric wave functions.

The atmosphere of weirdness and peculiarity surrounding quarks thickened into paradox when J. Friedman, H. Kendall, R. Taylor and their collaborators at the Stanford Linear Accelerator (SLAC) used energetic photons to poke into the inside of protons \cite{fkt}.  They discovered that there are indeed entities that look like quarks inside protons.   Surprisingly,  though, they found that when quarks are hit hard they seem to move (more accurately: to transport energy and momentum) as if they were free particles.   Before the experiment, most physicists had expected that whatever caused the strong interaction of quarks would also cause quarks to radiate energy abundantly, and thus rapidly to dissipate their motion, when they got violently accelerated.    

At a certain level of sophistication, this association of radiation with forces appears inevitable, and profound.  Indeed, the connection between forces and radiation is associated with some of the most glorious episodes in the history of physics.  In 1864 Maxwell predicted the existence of electromagnetic radiation -- including, but not limited to, ordinary light --  as a consequence of his consistent and comprehensive formulation of electric and magnetic forces.  Maxwell's new radiation was subsequently generated and detected by Hertz, in  1883 (and over the twentieth century its development has  revolutionized the way we manipulate matter and communicate with one another).   Much later, in 1935, Yukawa predicted the existence of pions based on his analysis of nuclear forces, and they were subsequently discovered in the late 1940s; the existences of many other hadrons were predicted successfully using a generalization these ideas.  (For experts: I have in mind the many resonances that were first seen in partial wave analyses, and then later in production.)  
More recently the existence of $W$ and $Z$ bosons, and of color gluons, and their properties, was inferred before their experimental discovery.  Those discoveries were, in 1972, still ahead of us, but they serve to confirm, retroactively, that our concerns were worthy ones.   Powerful interactions ought to be associated with powerful radiation. When the most powerful interaction in nature, the strong interaction, did not obey this rule, it posed a sharp paradox.

 \subsection{Paradox 2: Special Relativity and Quantum Mechanics Both Work}

The second paradox is more conceptual.   Quantum mechanics and special relativity are two great theories of twentieth-century physics.  Both are very successful.  But these two theories are based on entirely different ideas, which are not easy to reconcile.  In particular, special relativity puts space and time on the same footing, but quantum mechanics treats them very differently.  This leads to a creative tension, whose resolution has led to three previous Nobel Prizes (and ours is another).

The first of these prizes went to P. A. M. Dirac (1933).  
Imagine a particle moving on average at very nearly the speed of light, but with an uncertainty in position, as required by quantum theory.  Evidently it there will be some probability for observing this particle to move a little faster than average, and therefore faster than light, which special relativity won't permit.  The only known way to resolve this tension involves introducing the idea of antiparticles.  Very roughly speaking, the required uncertainty in position is accommodated by allowing for the possibility that the act of measurement can involve the creation of several  particles, each indistinguishable from the original, with different positions.   To maintain the balance of conserved quantum numbers, the extra particles must be accompanied by an equal number of antiparticles.    (Dirac was led to predict the existence of antiparticles through a sequence of ingenious interpretations and re-interpretations of the elegant relativistic wave equation he invented, rather than by heuristic reasoning of the sort I've presented.  The inevitability and generality of his conclusions, and their direct relationship to basic principles of quantum mechanics and special relativity, are only clear in retrospect.)  

The second and third of these prizes were to R. Feynman, J. Schwinger, and S.-I. Tomonaga (1965) and to G. 'tHooft and M. Veltman (1999) respectively.   The main problem that all these authors in one way or another addressed is the problem of ultraviolet divergences.

When special relativity is taken into account, quantum theory must allow for fluctuations in energy over brief intervals of time.  This is a generalization of the complementarity between momentum and position that is fundamental for ordinary, non-relativistic quantum mechanics.   Loosely speaking, energy can be borrowed to make evanescent virtual particles, including particle-antiparticle pairs.  Each pair passes away soon after it comes into being, but new pairs are constantly boiling up, to establish an equilibrium distribution.  In this way the wave function of (superficially) empty space becomes densely populated with virtual particles, and empty space comes to behave as a dynamical medium.

The virtual particles with very high energy create special problems.  If you calculate how much the properties of real particles and their interactions are changed by their interaction with virtual particles, you tend to get divergent answers, due to the contributions from virtual particles of very high energy.   

This problem is a direct descendant of the problem that triggered the introduction of quantum theory in the first place, i.e.  the ``ultraviolet catastrophe'' of black body radiation theory, addressed by Planck.  There the problem was that high-energy modes of the electromagnetic field are predicted, classically, to occur as thermal fluctuations, to such an extent that equilibrium at any finite temperature requires that there is an infinite amount of energy in these modes.  The difficulty came from the possibility of small-amplitude fluctuations with rapid variations in space and time.   The element of discreteness introduced by quantum theory eliminates the possibility of very small-amplitude fluctuations, because it imposes a lower bound on their size.   The (relatively) large-amplitude fluctuations that remain are predicted to occur very rarely in thermal equilibrium, and cause no problem.   
But quantum fluctuations are much more efficient than are thermal fluctuations at exciting the high-energy modes, in the form of virtual particles, and so those modes come back to haunt us.  For example, they give a divergent contribution to the energy of empty space, the so-called zero-point energy.

Renormalization theory was developed to deal with this sort of difficulty.  The central observation that is exploited in renormalization theory is that although interactions with high-energy virtual particles appear to produce divergent corrections, they do so in a very structured way.   That is, the same corrections appear over and over again in the calculations of many different physical processes.  For example in quantum electrodynamics (QED)  exactly two independent divergent expressions appear, one of which occurs when we calculate the correction to the mass of the electron, the other of which occurs when we calculate the correction to its charge.  To make the calculation mathematically well-defined, we must artificially exclude the highest energy modes, or dampen their interactions, a procedure called applying a cut-off, or regularization.  In the end we want to remove the cutoff, but at intermediate stages we need to leave it in, so as to have well-defined (finite) mathematical expressions.    If we are willing to take the mass and charge of the electron from experiment, we can identify the formal expressions for these quantities, including the potentially divergent corrections, with their measured values.   Having made this identification, we can remove the cutoff.  We thereby obtain well-defined answers, in terms of the measured mass and charge, for everything else of interest in QED.   

Feynman, Schwinger, and Tomonoga developed the technique for writing down the corrections due to interactions with any finite number of virtual particles in QED, and showed that renormalization theory worked in the simplest cases.   (I'm being a little sloppy in my terminology; instead of saying the number of virtual particles, it would be more proper to speak of the number of internal loops in a Feynman graph.)  Freeman Dyson supplied a general proof.  This was intricate work, that required new mathematical techniques.  'tHooft and Veltman showed that renormalization theory applied to a much wider class of theories, including the sort of spontaneously broken gauge theories that had been used by Glashow, Salam, and Weinberg to construct the (now) standard model of electroweak interactions.   Again, this was intricate and highly innovative work.

This brilliant work, however, still did not eliminate all the difficulties.  A very profound problem was identified by Landau \cite{landau}.   Landau argued that virtual particles would tend to accumulate around a real particle as long as there was any uncancelled influence.  This is called screening.  The only way for this screening process to terminate is for the source plus its cloud of virtual particles to cease to be of interest to additional virtual particles.  But then, in the end, no uncancelled influence would remain -- and no interaction!  
 
Thus all the brilliant work in QED and more general field theories represented, according to Landau, no more than a temporary fix.  You could get finite results for the effect of any particular number of virtual particles, but when you tried to sum the whole thing up, to allow for the possibility of an arbitrary number of virtual particles, you would get nonsense -- either infinite answers, or no interaction at all.  

Landau and his school backed up this intuition with calculations in many different quantum field theories.   They showed, in all the cases they calculated, that screening in fact occurred, and that it doomed any straightforward attempt to perform a complete, consistent calculation by adding up the contributions of more and more virtual particles.    We can sweep this problem under the rug in QED or in electroweak theory, because the answers including only a small finite number of virtual particles provide an excellent fit to experiment, and we make a virtue of necessity by stopping there.   But for the strong interaction that pragmatic approach seemed highly questionable, because there is no reason to expect that lots of virtual particles won't come into play, when they interact strongly.  

Landau thought that he had destroyed quantum field theory as a way of reconciling quantum mechanics and special relativity.  Something would have to give.   Either quantum mechanics or special relativity might ultimately fail, or else  essentially new methods would have to be invented, beyond quantum field theory, to reconcile them.  
Landau was not displeased with this conclusion, because in practice quantum field theory had not been very helpful in understanding the strong interaction, even though a lot of effort had been put into it.   But neither he, nor anyone else, proposed a useful alternative.

So we had the paradox, that combining quantum mechanics and special relativity seemed to lead inevitably to quantum field theory; but quantum field theory, despite substantial pragmatic success, self-destructed logically  due to catastrophic screening.

 \section{Paradox Lost: Antiscreening, or Asymptotic Freedom}
 
These paradoxes were resolved by our discovery of asymptotic freedom.  
 
We found that some very special quantum field theories actually have anti-screening.  We called this property asymptotic freedom, for reasons that will soon be clear.    Before describing the specifics of the theories, I'd like to indicate in a rough, general way how the phenomenon of antiscreening allows us to resolve our paradoxes.


Antiscreening turns Landau's problem on its head.  In the case of screening, a source of influence -- let us call it charge, understanding that it can represent something quite different from electric charge -- induces a canceling cloud of virtual particles.   From a large charge, at the center, you get a small observable influence far away.  
Antiscreening, or asymptotic freedom, implies instead that a charge of intrinsically small magnitude catalyzes a cloud of virtual particles that enhances its power.   I like to think of it as a thundercloud that grows thicker and thicker as you move away from the source.   

Since the virtual particles themselves carry charge, this growth is a self-reinforcing, runaway process.   
The situation appears to be out of control.  In particular, energy is required to build up the thundercloud, and the required energy threatens to diverge to infinity.   If that is the case, then the source could never be produced in the first place.  We've discovered a way to avoid Landau's disease -- by banishing the patients!  

At this point our first paradox, the confinement of quarks, makes a virtue of theoretical necessity.  For it suggests that there {\it are\/} in fact sources -- specifically, quarks -- that cannot exist on their own.   Nevertheless, Nature teaches us, these confined particles can play a role as building-blocks.   If we have, nearby to a source particle, its antiparticle (for example, quark and antiquark), then the catastrophic growth of the antiscreening thundercloud is no longer inevitable.   For where they overlap, the cloud of the source can be canceled by the anticloud of the antisource.    
Quarks and antiquarks, bound together, can be accommodated with finite energy, though either in isolation  would cause an infinite disturbance.  

Because it was closely tied to detailed, quantitative experiments, the sharpest problem we needed to address was the paradoxical failure of quarks to radiate when Friedman, Kendall, and Taylor subjected them to violent acceleration.   This too can be understood from the physics of antiscreening.   According to this mechanism, the color charge of a quark, viewed up close, is small.  It builds up its power to drive the strong interaction by accumulating a growing cloud at larger distances.   Since the power of its intrinsic color charge is small, the quark is actually only loosely attached to its cloud.    We can jerk it away from its cloud, and it will -- for a short while -- behave almost as if it had no color charge, and no strong interaction.  As the virtual particles in space respond to the altered situation they rebuild a new cloud, moving along with the quark, but this process does not involve significant radiation of energy and momentum.     That, according to us, was why you could analyze the most salient aspects of the SLAC experiments -- the inclusive cross-sections, which only keep track of overall energy-momentum flow -- as if the quarks were free particles, though in fact they are strongly interacting and ultimately confined.

Thus both our paradoxes, nicely dovetailed, get resolved together through antiscreening.


The theories that we found to display asymptotic freedom are called nonabelian gauge theories, or Yang-Mills theories \cite{ym}.  They form a vast generalization of electrodynamics.  They postulate the existence of several different kinds of charge, with complete symmetry among them.  So instead of one entity, ``charge'', we have several ``colors''.  Also, instead of one photon, we have a family of color gluons.   

The color gluons themselves carry color charges.  In this respect the nonabelian theories differ from electrodynamics, where the photon is electrically neutral.   Thus gluons in nonabelian theories play a much more active role in the dynamics of these theories than do photons in electrodynamics.   Indeed, it is the effect of virtual gluons that is responsible for antiscreening, which does not occur in QED.

It became evident to us very early on that one particular asymptotically free theory was uniquely suited as a candidate to provide {\it the\/} theory of the strong interaction.   On phenomenological grounds, we wanted to have the possibility to accommodate baryons, based on three quarks, as well as mesons, based on quark and antiquark.  In light of the preceding discussion, this requires that the color charges of three different quarks can cancel, when you add them up.   This can oocur if the three colors exhaust all possibilities; so we arrived at the gauge group $SU(3)$, with three colors, and eight gluons.  To be fair, several physicists had, with various motivations, suggested the existence of a three-valued internal color label for quarks years before \cite{nambu}.   It did not require a great leap of imagination to see how we could adapt those ideas to our tight requirements.


By using elaborate technical machinery of quantum field theory (including the renormalization group, operator product expansions, and appropriate dispersion relations) we were able to be much more specific and quantitative about the implications our theory than my loose pictorial language suggests.  In particular, the strong interaction does not simply turn off abruptly, and there is a non-zero probability that quarks will radiate when poked.  It is only asymptotically, as energies involved go to infinity, that the probability for radiation vanishes.   We could calculate in great detail the observable effects of the radiation at finite energy,  and make experimental predictions based on these calculations.   At the time, and for several years later, the data was not accurate enough to test these particular predictions, but by the end of the 1970s they began to look good, and by now they're beautiful.

Our discovery of asymptotic freedom, and its essentially unique realization in quantum field theory, led us to a new attitude towards the problem of the strong interaction.  
In place of the broad research programs and fragmentary insights that had characterized earlier work, we now had a single, specific candidate theory -- a theory that could be tested, and perhaps falsified, but which could not be fudged.  Even now, when I re-read our declaration \cite{gw2}
\begin{quotation}
Finally let us recall that the proposed theories appear to be uniquely singled out by nature, if one takes both the SLAC results and the renormalization-group approach to quantum field theory at face value.
\end{quotation}
I re-live the mixture of exhilaration and anxiety that I felt at the time. 

 \section{A Foursome of Paradigms}
 
Our resolution of the paradoxes that drove us had ramifications in unanticipated directions, and extending far beyond their initial scope.

 \subsection{Paradigm 1: The Hard Reality of Quarks and Gluons}

Because, in order to fit the facts,  you had to ascribe several bizarre properties to quarks -- paradoxical dynamics, peculiar charge, and anomalous statistics -- their ``reality'' was, in 1972, still very much in question.  This despite the fact that they were helpful in organizing the hadrons, and even though Friedman, Kendall, and Taylor had ``observed'' them!    The experimental facts wouldn't go away, of course, but their ultimate significance remained doubtful.  Were quarks  basic particles, with simple properties, that could be used to in formulating a profound theory -- or just a curious intermediate device, that would need to be replaced by deeper conceptions?  

Now we know how the story played out, and it requires an act of imagination to conceive how it might have been different.  But Nature is imaginative, and so are theoretical physicists, and so  it's not impossible to fantasize alternative histories.  For example, the quasiparticles of the fractional quantum Hall effect, which are not basic but rather emerge as collective excitations involving ordinary electrons, also cannot exist in isolation, and they have fractional charge and anomalous statistics!    Related things happen in the Skyrme model, where nucleons emerge as collective excitations of pions.    One might have fantasized that quarks would follow a similar script, emerging somehow as collective excitations of hadrons, or of more fundamental preons, or of strings.   

Together with the new attitude toward the strong interaction problem, that I just mentioned, came a new attitude toward
quarks and gluons.  These words were no longer just names attached to empirical patterns, or to notional building blocks within rough phenomenological models.  Quarks and (especially) gluons had become ideally simple entities, whose properties are fully defined by mathematically precise algorithms.  

You can even see them! Here's a picture, which I'll now explain.
\begin{figure}[ht] 
   \centering
   \includegraphics[width=3in]{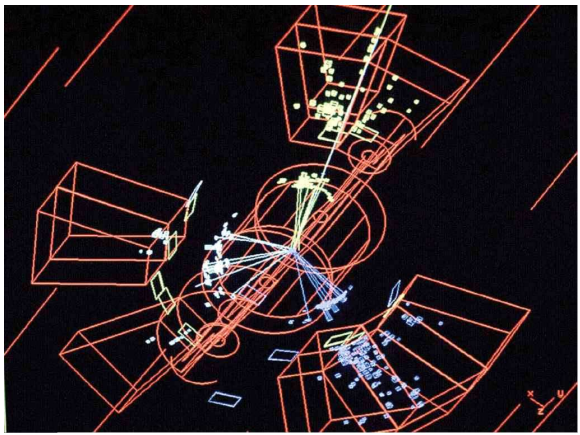} 
  \caption{A photograph from the L3 collaboration, showing three jets emerging from electron-positron annihilation at  high energy \cite{L3}.  These jets are the materialization of a quark, antiquark, and gluon.}
  \label{fig:1}
\end{figure}

Asymptotic freedom is a great boon for experimental physics, because it leads to the beautiful phenomenon of jets.
As I remarked before, an important part of the atmosphere of mystery surrounding quarks arose from the fact that they could not be isolated.    
But if we change our focus, to follow flows of energy and momentum rather than individual hadrons, then quarks and gluons come into view, as I'll now explain. 

There is a contrast between two different kinds of radiation, which expresses the essence of asymptotic freedom.  
Hard radiation, capable of significantly re-directing the flow of energy and momentum,  is rare.   But soft radiation, that produces additional particles moving in the same direction, without deflecting the overall flow,  is common.  Indeed, soft radiation is associated with the  build-up of the clouds I discussed before, as it occurs in time.    Let's consider what it means for experiments, say to be concrete the sort of experiment done at the Large Electron Positron collider (LEP) at CERN during the 1990s, and contemplated for the International Linear Collider (ILC) in the future.  At these facilities, one studies what emerges from the annihilation of electrons and positrons that collide at high energies. 
\begin{figure}[t] 
   \centering
   \includegraphics[scale=.75]{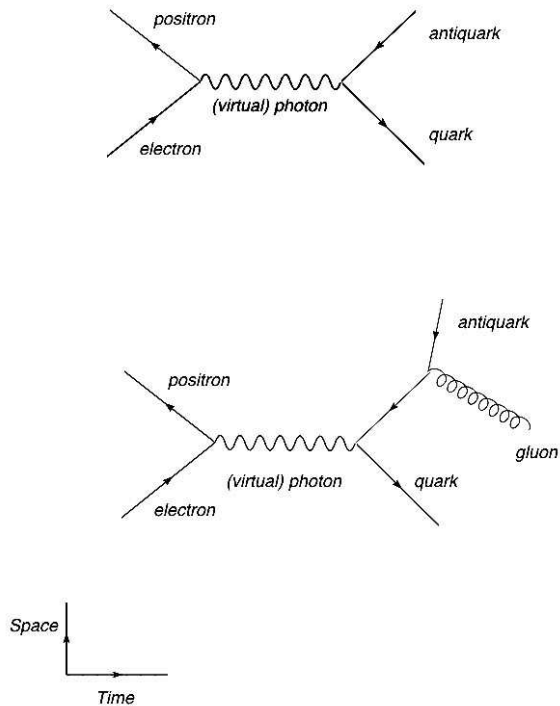} 
   \caption{These Feynman graphs are schematic representations of the fundamental processes in electron-positron annihilation, as they take place in space and time.  They show the origin of two-jet and three-jet events. }
   \label{fig:2}
\end{figure}
By well-understood processes that belong to QED or electroweak theory, the annihilation proceeds through a virtual photon or $Z$ boson into a quark and an antiquark.   Conservation and energy and momentum dictate that the quark and antiquark will be moving at high speed in opposite directions.  If there is no hard radiation, then the effect of soft radiation will be to convert the quark into a spray of hadrons moving in a common direction: a jet.   Similarly, the antiquark becomes a jet moving in the opposite direction.  The observed result is then a 2-jet event.   Occasionally (about 10\%  of the time, at LEP) there will be hard radiation, with the quark (or antiquark) emitting a gluon in a significantly new direction.   From that point on the same logic applies, and we  have a 3-jet event, like the one shown in  Figure \ref{fig:1}.  The theory of the underlying space-time process is depicted in Figure \ref{fig:2}.  And roughly 1\% of the time 4 jets will occur, and so forth.   The relative probability of different numbers of jets, how it varies with the overall energy, the relative frequency of different angles at which the jets emerge and the total energy in each -- all these detailed aspects of the ``antenna pattern'' can be predicted quantitatively.  These predictions reflect the basic couplings among quarks and gluons, which define QCD, quite directly.   

The predictions agree well with very comprehensive experimental measurements.  So we can conclude with confidence that QCD is right, and that what you are seeing, in Figure \ref{fig:1}, is a quark, an antiquark, and a gluon -- although, since the predictions are statistical, we can't say for sure which is which!     

\begin{figure}[h] 
   \centering 
   \includegraphics[width=3.15in]{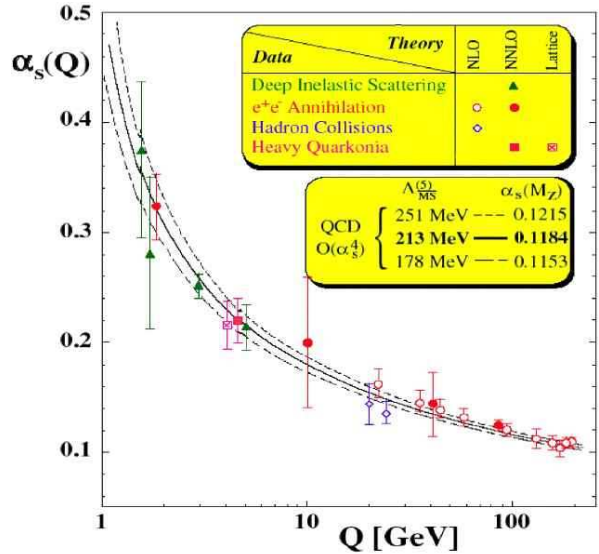}
   \caption{Many quite different experiments, performed at different energies, have been successfully analyzed using QCD.  Each fits a large quantity of data to a single parameter, the strong coupling $\alpha_s$. By comparing the values they report, we obtain direct confirmation that the coupling evolves as predicted \cite{bethke}. }
   \label{fig:3}
\end{figure}

By exploiting the idea that hard radiation processes, reflecting fundamental quark and gluon interactions, control the overall flow of energy and momentum in high-energy processes, one can analyze and predict the behavior of many different kinds of experiments.  In most of these applications, including the original one to deep inelastic scattering, the analysis necessary to separate out hard and soft radiation is much more involved and harder to visualize than in the case of electron-positron annihilation.  A lot of ingenuity has gone, and continues to go, into this subject, known as perturbative QCD. The results have been quite successful and gratifying.  Figure \ref{fig:3} shows one aspect of the success.  Many different kinds of experiments, performed at many different energies, have been successfully described by QCD predictions, each in terms of the one relevant parameter of the theory, the overall coupling strength.  Not only must each experiment, which may involve hundreds of independent measurements, be fit consistently, but one can then check whether the values of the coupling change with the energy scale in the way we predicted.  As you can see, it does.  A remarkable tribute to the success of the theory, which I've been amused to watch evolve, is that a lot of the same activity that used to be called {\it testing QCD\/} is now called {\it calculating backgrounds}.  

As a result of all this success, a new paradigm has emerged for the operational meaning of the concept of a fundamental particle.  Physicists designing and interpreting high-energy experiments now routinely describe their results in terms of producing and detecting quarks and gluons: what they mean, of course, is the corresponding jets.   

 \subsection{Paradigm 2: Mass Comes from Energy}

My friend and mentor Sam Treiman liked to relate his experience of how, during World War II, the U.S. Army responded to the challenge of training a large number of radio engineers starting with very different levels of preparation, ranging down to near zero.  They designed a crash course for it, which Sam took.  In the training manual, the first chapter was devoted to Ohm's three laws.  
Ohm's first law is $V=IR$.  Ohm's second law is $I=V/R$.  I'll leave it to you to reconstruct Ohm's third law.

Similarly, as a companion to Einstein's famous equation $E=mc^2$ we have his second law, $m=E/c^2$.  

All this isn't quite as silly as it may seem, because different forms of the same equation can suggest very different things.  The usual way of writing the equation, $E=mc^2$, suggests the possibility of obtaining large amounts of energy by converting small amounts of mass.  It brings to mind the possibilities of nuclear reactors, or bombs.   
Stated as  $m=E/c^2$, Einstein's law suggests the possibility of explaining mass in terms of energy.  That is a good thing to do, because in modern physics energy is a more basic concept than mass.   
Actually, Einstein's original paper does not contain the equation $E=mc^2$, but rather $m=E/c^2$.  In fact, the title is a question: ``Does the Inertia of a Body Depend Upon its Energy Content?'' From the beginning, Einstein was thinking about the origin of mass, not about making bombs.  

Modern QCD answers Einstein's question with a resounding ``Yes!''  Indeed, the mass of ordinary matter derives almost entirely from energy -- the energy of massless gluons and nearly massless quarks,  which are the ingredients from which protons, neutrons, and atomic nuclei are made.   

The runaway build-up of antiscreening clouds, which I described before, cannot continue indefinitely.   The resulting color fields would carry infinite energy, which is not available.  The color charge that threatens to induce this runaway must be cancelled.   The color charge of a quark can be cancelled either with an antiquark of the opposite color (making a meson), or with two quarks of the complementary colors (making a baryon).  In either case, perfect cancellation would occur only if the particles doing the canceling were located right on top of the original quark -- then there would be no uncanceled source of color charge anywhere in space, and hence no color field.   Quantum mechanics does not permit this perfect cancellation, however.   The quarks and antiquarks are described by wave functions, and spatial gradients in these wave function cost energy, and so there is a high price to pay for localizing the wave function within a small region of space.  Thus, in seeking to minimize the energy, there are two conflicting considerations: to minimize the field energy, you want to cancel the sources accurately; but to minimize the wave-function localization energy, you want to keep the sources fuzzy.  The stable configurations will be based on different ways of compromising between these two considerations.   In each such configuration, there will be both field energy and localization energy.  This gives rise to mass, according to $m=E/c^2$, even if the gluons and quarks started out without any non-zero mass of their own.   So the different stable compromises will be associated with particles that we can observe, with different masses; and metastable compromises will be associated with observable particles that have finite lifetimes.

To determine the stable compromises concretely, and so to predict the masses of mesons and baryons, is hard work.  It 
requires difficult calculations that continue to push the frontiers of massively parallel processing. I find it quite ironical that if we want to compute the mass of a proton, we need to deploy something like $10^{30}$ protons and neutrons, doing trillions of multiplications per second, working for months, to do what one proton does in $10^{-24}$ seconds, namely figure out its mass.   Maybe it qualifies as a paradox.  At the least, it suggests that there may be much more efficient ways to calculate than the ones we're using.     

\begin{figure}[h] 
    \centering
    \includegraphics[width=4in]{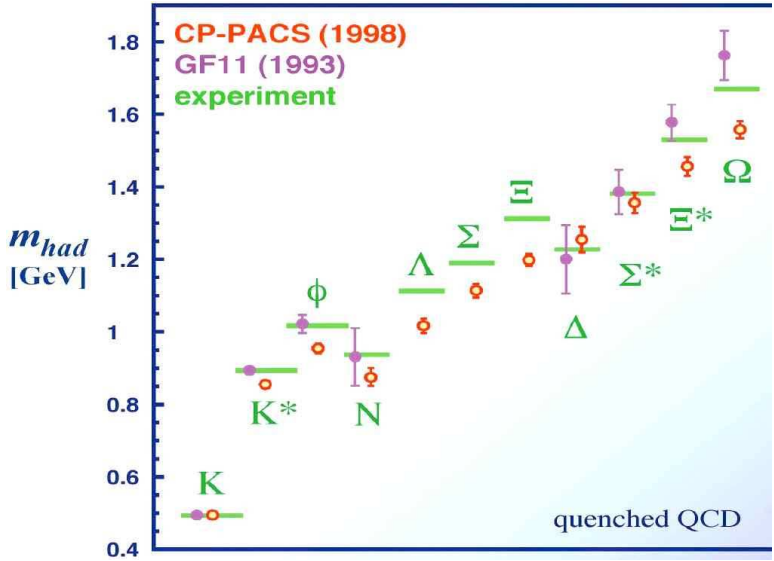} 
    \caption{Comparison of observed hadron masses to the energy spectrum predicted by QCD, upon direct numerical integration of the equations, exploiting immense computer power \cite{lattice}.  The small remaining discrepancies are consistent with what is expected given the approximations that were necessary to make the calculation practical.}
    \label{fig:4}
 \end{figure}

In any case, the results that emerge from these calculations are very gratifying.  They are displayed in \ref{fig:4}.
The observed masses of prominent mesons and baryons are reproduced quite well, stating from an extremely tight and rigid theory.  Now is the time to notice also that one of the data points in Figure \ref{fig:3}, the one labeled ``Lattice'', is of a quite different character from the others.  It is based not on the perturbative physics of hard radiation, but rather on the comparison of a direct integration of the full equations of QCD with experiment, using the techniques of lattice gauge theory.  

The success of these calculations represents the ultimate triumph over our two paradoxes:
\begin{itemize}
\item The calculated spectrum does not contain anything with the charges or other quantum numbers of quarks; nor of course does it contain massless gluons.  The observed particles do not map in a straightforward way to the primary fields from which they ultimately arise.  
\item Lattice discretization of the quantum field theory provides a cutoff procedure that is independent of any expansion in the number of virtual particle loops.  The renormalization procedure must be, and is, carried out without reference to perturbation theory, as one takes the lattice spacing to zero.  Asymptotic freedom is crucial for this, as I discussed -- it saves us from Landau's catastrophe.  
\end{itemize}

By fitting some fine details of the pattern of masses, one can get an estimate of what the quark masses are, and how much their masses are contributing to the mass of the proton and neutron.  It turns out that what I call QCD Lite --  the version in which you put the $u$ and $d$ quark masses to zero, and ignore the other quarks entirely -- provides a remarkably good approximation to reality.  Since QCD Lite is a theory whose basic building-blocks have zero mass, this result quantifies and makes precise the idea that most of the mass of ordinary matter -- 90 \% or more -- arises from pure energy, via $m=E/c^2$.

The calculations make beautiful images, if we work to put them in eye-friendly form. 
Derek Leinweber has done made some striking  animations of QCD fields as they fluctuate in empty space.  Figure \ref{fig:DerekL}  is a snapshot from one of his animations.   Figure \ref{fig:GregK}  from Greg Kilcup, displays the (average) color fields, over and above the fluctuations, that are  associated with a very simple hadron, the pion, moving through space-time.  Insertion of a quark-antiquark pair, which we subsequently remove, produces this disturbance in the fields.
\begin{figure}[h] 
   \centering
   \begin{minipage}[c]{0.4\textwidth}
   \centering
   \includegraphics[width=2in]{newfigure5.epsf} 
   \caption{A snapshot of spontaneous quantum fluctuations in the gluon fields \cite{leinweber}.  For experts: what is shown is the topological charge density in a typical contribution to the functional integral, with high-frequency modes filtered out.}
   \label{fig:DerekL}
   \vspace{.90in}
   \end{minipage}
   \qquad
   \begin{minipage}[c]{0.4\textwidth}
   \centering
   \includegraphics[width=2in]{newfigure6.epsf} 
  \caption{The calculated net distribution of field energy caused by injecting and removing a quark-antiquark pair \cite{kilcup}.    By calculating the energy in these fields, and the energy in analogous fields produced by other disturbances, we predict the masses of hadrons.  In a profound sense, these fields {\it are\/} the hadrons.}
   \label{fig:GregK}
   \end{minipage}
\end{figure}

These pictures make it clear and tangible that the quantum vacuum is a dynamic medium, whose properties and responses largely determine the behavior of matter.   In quantum mechanics, energies are associated with frequencies, according to the Planck relation $E=h\nu$.  The masses of hadrons, then, are uniquely associated to tones emitted by the dynamic medium of space when it disturbed in various ways, according to 
\begin{equation}
\nu = mc^2/h
\end{equation}
We thereby discover, in the reality of masses, an algorithmic, precise Music of the Void.  It is a modern embodiment of the ancients' elusive, mystical ``Music of the Spheres''.

\subsection{Paradigm 3: The Early Universe was Simple}

In 1972 the early universe seemed hopelessly opaque.  In conditions of ultra-high temperatures, as occurred close to the Big Bang singularity, one would have lots of hadrons and antihadrons, each one an extended entity that interacts strongly and in complicated ways with its neighbors.   They'd start to overlap with one another, and thereby produce a theoretically intractable mess.   

But asymptotic freedom renders ultra-high temperatures friendly to theorists.  It says that if we switch from a description based on hadrons to a description based on quark and gluon variables, and focus on quantities like total energy, that are not sensitive to soft radiation, then the treatment of the strong interaction, which was the great difficulty, becomes simple.  We can calculate to a first approximate by pretending that the quarks, antiquarks and gluons behave as free particles, then add in the effects of rare hard interactions.   This makes it quite practical to formulate a precise description of the properties of ultra-high temperature matter that are relevant to cosmology.  
  
\begin{figure}[ht]
 \centering
\includegraphics[scale=.30]{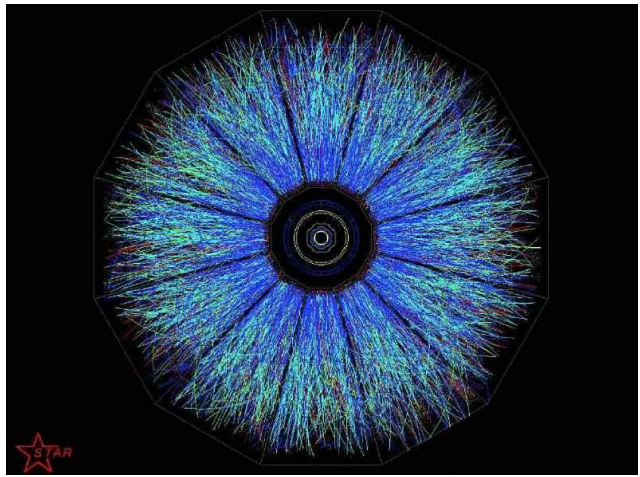}
\caption{A picture of particle tracks emerging from the collision of two gold ions at high energy.  The resulting fireball and its subsequent expansion recreate, on a small scale and briefly, physical conditions that last occurred during the Big Bang \cite{STAR}.}
\label{fig:7}
\end{figure}

We can even, over an extremely limited volume of space and time, reproduce Big Bang conditions in terrestrial laboratories.  When heavy ions are caused to collide at high energy, they produce a fireball that briefly attains temperatures as high as 200 MeV.  ``Simple'' may not be the word that occurs to you in describing the explosive outcome of this event, as displayed in Figure \ref{fig:7}, but in fact detailed study does permit us to reconstruct aspects of the initial fireball, and to check that it was a plasma of quarks and gluons.

 \subsection{Paradigm 4: Symmetry Rules}

Over the course of the twentieth century, symmetry has been immensely fruitful as a source of insight into Nature's basic operating principles.   QCD, in particular, is constructed as the unique embodiment of a huge symmetry group, local $SU(3)$ color gauge symmetry (working together with special relativity, in the context of quantum field theory).   As we try to discover new laws, that improve on what we know, it seems good strategy to continue to use symmetry as our guide.    This strategy has led physicists to several compelling suggestions, which I'm sure you'll be hearing more about in future years!   QCD plays an important role in all of them -- either directly, as their inspiration, or as an essential tool in devising strategies for experimental exploration.  

I will discuss one of these suggestions schematically, and mention three others telegraphically.  
 
 \subsubsection{Unified Field Theories}

Both QCD and the standard electroweak standard model are founded on gauge symmetries.  This combination of theories gives a wonderfully economical and powerful account of an astonishing range of phenomena.  Just because it is so concrete and so successful, this rendering of Nature can and should be closely scrutinized for its aesthetic flaws and possibilities.   Indeed, the structure of the gauge system gives powerful suggestions for
its further fruitful development. Its product structure
$SU(3)\times SU(2) \times U(1)$, the reducibility of the fermion representation
(that is, the fact that the symmetry does not make connections linking
all the fermions), and the peculiar values of the quantum number
hypercharge assigned to the known particles all suggest the desirability of a larger symmetry.   

The devil is in the details, and it is not at all automatic that the
superficially complex and messy observed pattern of matter will fit neatly into a simple
mathematical structure. But, to a remarkable extent, it does.
\begin{figure}[htbp] 
    \centering
    \includegraphics[width=4in]{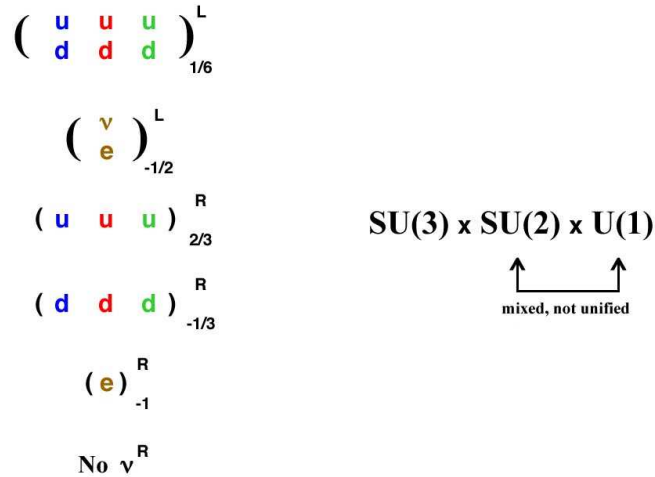} 
    \caption{A schematic representation of the symmetry structure of the standard model.  There are three independent symmetry transformations, under which the known fermions fall into five independent units (or fifteen, after threefold family repetition).  The color gauge group $SU(3)$ of QCD acts horizontally, the weak interaction gauge group $SU(2)$ acts vertically, and the hypercharge $U(1)$ acts with the relative strengths indicated by the subscripts.  Right-handed neutrinos do not participate in any of these symmetries.}
    \label{fig:8}
 \end{figure}
 
Most of what we know about the strong, electromagnetic, and weak interactions is summarized (rather schematically!) in Figure \ref{fig:8}.  QCD connects particles horizontally in groups of 3 ($SU(3)$), the weak interaction connects particles vertically in groups of 2 ($SU(2)$)  in the horizontal direction and hypercharge ($U(1)$) senses the little subscript numbers.  Neither the different interactions, nor the different particles, are unified.  There are three different interaction symmetries, and five disconnected sets of particles (actually fifteen sets, taking into account the threefold repetition of families).

\begin{figure}[htbp] 
    \centering
    \includegraphics[width=3.75in]{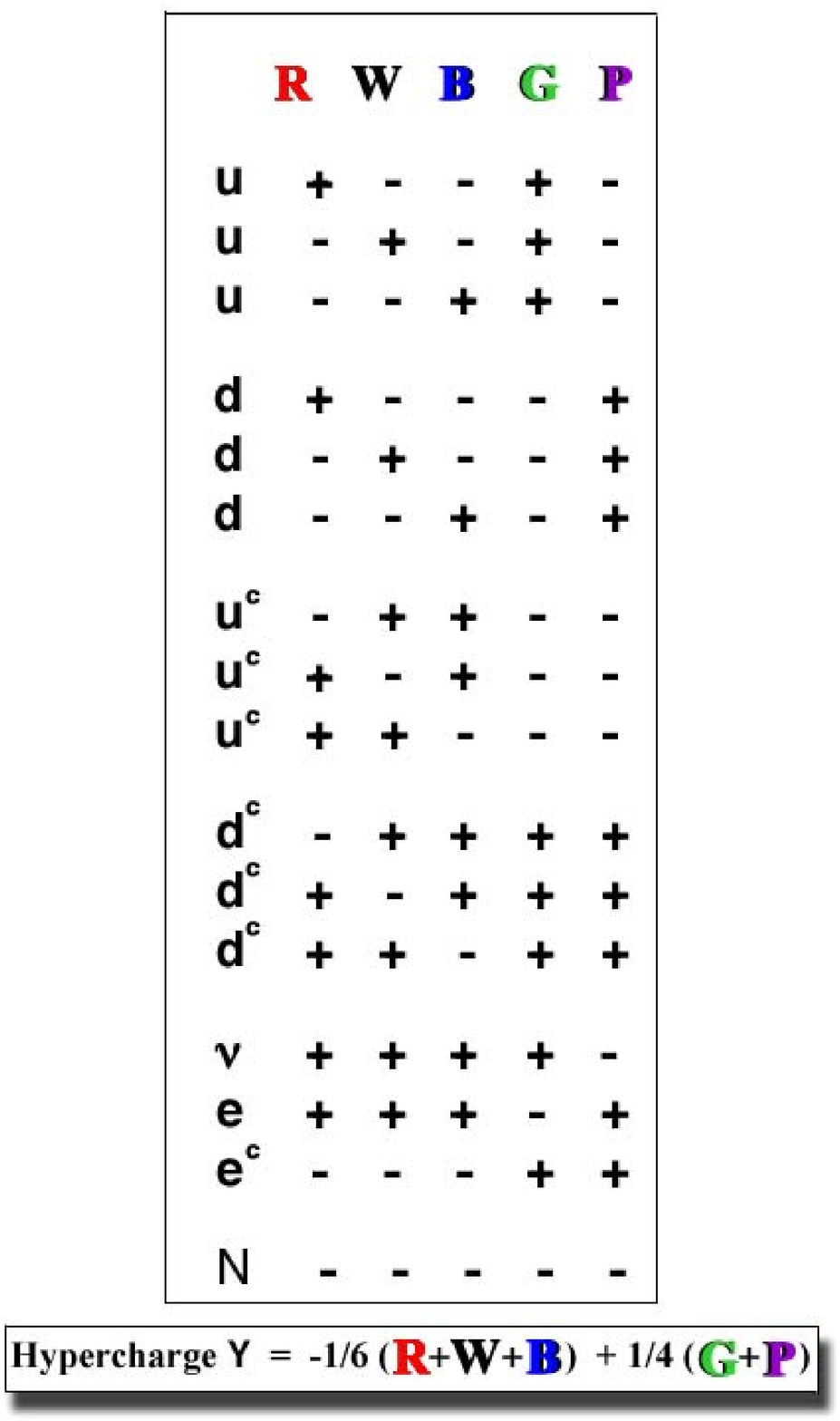} 
    \caption{The hypothetical enlarged symmetry $SO(10)$ \cite{georgi} accommodates all the symmetries of the standard model, and more, into a unified mathematical structure.   The fermions, including a right-handed neutrino that plays an important role in understanding observed neutrino phenomena, now form an irreducible unit (neglecting family repetition).   The allowed color charges, both strong and weak, form a perfect match to what is observed.  The phenomenologically required hypercharges, which appear so peculiar in the standard model, are now theoretically determined by the color and weak charges, according to the formula displayed.}
    \label{fig:9}
 \end{figure}

We can do much better by having more symmetry, implemented by additional gluons that also change strong into weak colors.   Then everything clicks into place quite beautifully, as displayed in Figure \ref{fig:9}.   
  
\begin{figure}[htbp] 
    \centering
    \includegraphics[width=4in]{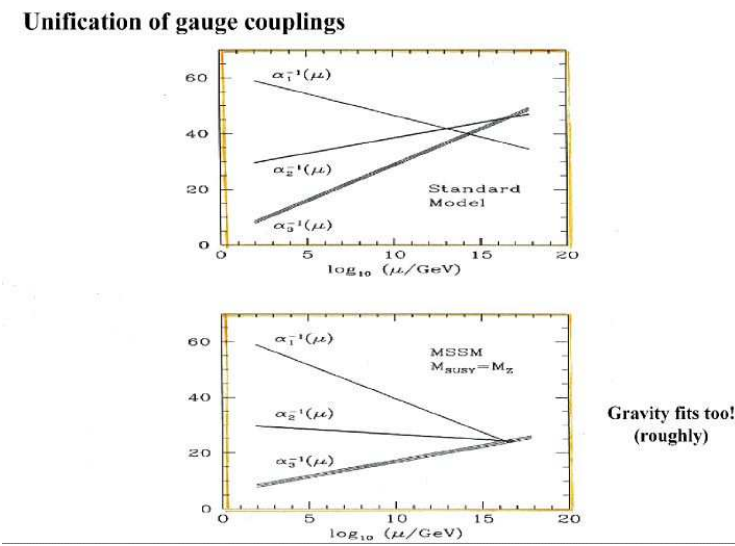} 
    \caption{We can test the hypothesis that the disparate coupling strengths of the different gauge interactions derive a common value at short distances, by doing calculations to take into account the effect of virtual particle clouds \cite{gqw}.  These are the same sort of calculations that go into Figure \ref{fig:3}, but extrapolated to much higher energies, or equivalently shorter distances.  Top panel: using known virtual particles.  Bottom panel: including also the virtual particles required by low-energy  supersymmetry \cite{drw}.}
    \label{fig:10}
 \end{figure}

There seems to be a problem, however.   The different interactions, as observed, do not have the same overall strength, as would be required by the extended symmetry.   Fortunately, asymptotic freedom informs us that the observed interaction strengths at a large distance can be different from the basic strengths of the seed couplings viewed at short distance.   To see if the basic theory might have the full symmetry, we have to look inside the clouds of virtual particles, and to track the evolution of the couplings.  We can do this, using the same sort of calculations that underlie Figure \ref{fig:3}, extended to include the electroweak interactions, and extrapolated to much shorter distances (or equivalently, larger energy scales).   It is convenient to display inverse couplings and work on a logarthmic scale, for then the evolution is (approximately) linear.  
When we do the calculation using only the virtual particles for which we have convincing evidence, we find that the couplings do approach each other in a promising way, though ultimately they don't quite meet This is shown in the top panel of Figure \ref{fig:10}.   

Interpreting things optimistically, we might surmise from this near-success that the general idea of unification is on the right track, as is our continued reliance on quantum field theory to calculate the evolution of couplings.   After all, it is hardly shocking that  extrapolation of the equations for evolution of the couplings beyond their observational foundation by many orders of magnitude is missing some quantitatively significant ingredient.   In a moment I'll mention an attractive hypothesis for what's missing.   

A very general consequence of this line of thought is that an enormously large energy scale, of order $10^{15}$ GeV  or more, emerges naturally as the scale of unification. This is a profound and welcome result.  It is profound, because the large energy scale -- which is far beyond any energy we can access directly -- emerges from careful consideration of experimental realities at energies more than ten orders of magnitude smaller!  The underlying logic that gives us this leverage is a synergy of unification and asymptotic freedom, as follows.  If evolution of couplings is to be responsible for their observed gross inequality then, since this evolution is only logarithmic in energy, it must act over a very wide range.  

The emergence of a large mass scale for unification is welcome, first, because many effects we might expect to be associated with unification are observed to be highly suppressed.   Symmetries that unify $SU(3)\times SU(2) \times U(1)$ will almost inevitably involve wide possibilities for transformation among quarks, leptons, and their antiparticles.   These extended possibilities of transformation, mediated by the corresponding gauge bosons, undermine conservation laws including lepton and baryon number conservation.   Violation of lepton number is closely associated with neutrino oscillations.  Violation of baryon number is closely associated with proton instability.    In recent years neutrino oscillations have been observed; they correspond to miniscule neutrino masses, indicating a very feeble violation of lepton number.   Proton instability has not yet been observed, despite heroic efforts to do so.  In order to keep these processes sufficiently small, so as to be consistent with observation, a high scale for unification, which suppresses the occurrence of the transformative gauge bosons as virtual particles, is most welcome.   In fact, the unification scale we infer from the evolution of couplings is broadly consistent with the observed value of neutrino masses and encourages further vigorous pursuit of the quest to observe proton decay.   
 
The emergence of a large mass scale for unification is welcome, secondly, because it opens up possibilities for making quantitative connections to the remaining fundamental interaction in Nature: gravity.    It is notorious that gravity is absurdly feebler than the other interactions, when they are compared acting between fundamental particles at accessible energies.  The gravitational force between proton and electron, at any macroscopic distance, is about $G m_e m_p / \alpha \sim 10^{-40}$ of the electric force.   On the face of it, this fact poses a severe challenge to the idea that these forces are different manifestations of a common source -- and an even more severe challenge to the idea that gravity, because of its deep connection to space-time dynamics, is the primary force.   

By extending our consideration of the evolution of couplings to include gravity, we can begin to meet these challenges.
\begin{itemize}
\item 
Whereas the evolution of gauge theory couplings with energy is a subtle quantum-mechanical effect, the gravitational coupling evolves even classically, and much more rapidly.  For gravity responds directly to energy-momentum, and so it appears stronger when viewed with high-energy probes.  In moving from the small energies where we ordinarily measure to unification energy scales, the ratio $G E^2 / \alpha$ ascends to values that are no longer absurdly small.  
\item
If gravity is the primary force, and special relativity and quantum mechanics frame the discussion, then Planck's system of physical units, based on Newton's constant $G$, the speed of light $c$, and Planck's quantum of action $h$, is privileged.  Dimensional analysis then suggests that the value of naturally defined quantities, measured in these units, should be of order unity.  But when we measure the proton mass in Planck units, we discover 
\begin{equation}
m_p \sim 10^{-18} \sqrt {\frac{hc}{G}}
\end{equation}
On this hypothesis, it makes no sense to ask ``Why is gravity so feeble?''.  Gravity, as the primary force, just is what it is.   The right question is the one we confront here: ``Why is the proton so light?''.   Given our new, profound understanding of the origin of the proton's mass, which I've sketched for you today, we can formulate a tentative answer.  The proton's mass is set by the scale at which the strong coupling, evolved down from its primary value at the Planck energy, comes to be of order unity.  It is then that it becomes worthwhile to cancel off the growing color  fields of quarks, absorbing the cost of quantum localization energy.   In this way, we find, quantitatively, that the tiny value of the proton mass in Planck units arises from the fact that the basic unit of color coupling strength, $g$, is of order $\frac{1}{2}$ at the Planck scale!  Thus dimensional reasoning is no longer mocked.  The apparent feebleness of gravity results from our partiality toward the perspective supplied by matter made from protons and neutrons.  
\end{itemize}

\subsubsection{Supersymmetry}
 
As I mentioned a moment ago, the approach of couplings to a unified value is suggested, but not accurately realized, if we infer their evolution by including the effect of known virtual particles.  There is one particular proposal to expand the world of virtual particles, which is well motivated on several independent grounds. It is known as low-energy supersymmetry \cite{nilles}.

As the name suggests, supersymmetry involves expanding the symmetry of the basic equations of physics.  This proposed expansion of symmetry goes in a different direction from the enlargement of gauge symmetry.   Supersymmetry  makes transformations between particles having the same color charges and different spins, whereas expanded gauge symmetry changes the color charges while leaving spin untouched.  Supersymmetry expands the space-time symmetry of special relativity.   

In order to implement low-energy supersymmetry, we must postulate the existence of a whole new world of heavy particles, none of which has yet been observed directly.  There is, however, a most intriguing indirect hint that this idea may be on the right track: If we include the particles needed for low-energy supersymmetry, in their virtual form, in the calculation of how couplings evolve with energy, then accurate unification is achieved!    This is shown in the bottom panel of Figure \ref{fig:10}.

By ascending a tower of speculation, involving now both extended gauge symmetry and extended space-time symmetry, we seem to break though the clouds, into clarity and breathtaking vision.  Is it an illusion, or reality?  This question creates a most exciting situation for the Large Hadron Collider (LHC), due to begin operating at CERN in 2007, for this great accelerator will achieve the energies necessary to access the new world of of heavy particles, if it exists.  How the story will play out, only time will tell.  But in any case I think it is fair to say that the pursuit of unified field theories, which in past (and many present) incarnations has been vague and not fruitful of testable consequences, has in the circle of ideas I've been describing here attained entirely new levels of concreteness and fecundity.  
 
 \subsubsection{Axions \cite{axions}}

As I have emphasized repeatedly, QCD is in a profound and literal sense constructed as the embodiment of symmetry.   There is an almost perfect match between the observed properties of quarks and gluons and the most general properties allowed by color gauge symmetry, in the framework of special relativity and quantum mechanics.   The exception is that the established symmetries of QCD fail to forbid one sort of behavior that is not observed to occur.  The established symmetries permit a sort of interaction among gluons -- the so-called $\theta$ term -- that violates the invariance of the equations of QCD under a change in the direction of time.   Experiments provide extremely severe limits on the strength of this interaction, much more severe than might be expected to arise accidentally.  

By postulating a new symmetry, we can explain the absence of the undesired interaction.  The required symmetry is called Peccei-Quinn symmetry after the physicists who first proposed it.  If it is present, this symmetry has remarkable consequences.   It leads us to predict the existence of new very light, very weakly interacting particles, {\it axions}.     (I named them after a laundry detergent, since they clean up a problem with an axial current.)  In principle axions might be observed in a variety of ways, though none is easy.   They have interesting implications for cosmology, and they are a leading candidate to provide cosmological dark matter.   
 
 \subsubsection{In Search of Symmetry Lost \cite{symmLost}}

It has been almost four decades since our current, wonderfully
successful theory of the electroweak interaction
was formulated. Central to that theory is the
concept of spontaneously broken gauge symmetry.
According to this concept, the fundamental equations
of physics have more symmetry than the actual physical
world does. Although its specific use in electroweak theory
involves exotic hypothetical substances and some sophisticated
mathematics, the underlying theme of broken
symmetry is quite old. It goes back at least to the dawn of
modern physics, when Newton postulated that the basic laws
of mechanics exhibit full symmetry in three dimensions of
space Ñ despite the fact that everyday experience clearly distinguishes
`up and down' from `sideways' directions in our
local environment. Newton, of course, traced this asymmetry
to the influence of Earth's gravity. In the framework of
electroweak theory, modern physicists similarly postulate
that the physical world is described by a solution wherein all
space, throughout the currently observed Universe, is permeated
by one or more (quantum) fields that spoil the full
symmetry of the primary equations. 

Fortunately this hypothesis, which might at first hearing
sound quite extravagant, has testable implications. The
symmetry-breaking fields, when suitably excited, must bring
forth characteristic particles: their quanta. Using the most economical
implementation of the required symmetry breaking,
one predicts the existence of a remarkable new particle, the so-called
Higgs particle. More ambitious speculations suggest
that there should be not just a single Higgs particle, but rather a
complex of related particles.  Low-energy supersymmetry, for example, requires at least five ``Higgs particles''.

Elucidation of the Higgs complex will be another major task for the LHC.  In planning this endeavor, QCD and asymptotic freedom play a vital supporting role.   The strong interaction will be responsible for most of what occurs in collisions at the LHC.  To discern the new effects, which will be manifest only in a small proportion of the events, we must understand the dominant backgrounds very well.   Also, the production and decay of the Higgs particles themselves usually involves quarks and gluons.  To anticipate their signatures, and eventually to interpret the observations, we must use our understanding of how protons -- the projectiles at LHC -- are assembled from quarks and gluons, and how quarks and gluons show themselves as jets.

 \section{The Greatest Lesson}

Evidently asymptotic freedom, besides resolving the paradoxes that originally concerned us, provides a conceptual foundation for several major insights into Nature's fundamental workings, and a versatile instrument for further investigation.

The greatest lesson, however, is a moral and philosophical one.  It is truly awesome to discover, by example, that we humans can come to comprehend Nature's deepest principles, even when they are  hidden in remote and alien realms.  Our minds were not created for this task, nor were appropriate tools ready at hand.  Understanding was achieved through a vast international effort involving thousands of people working hard for decades, competing in the small but cooperating in the large, abiding by rules of openness and honesty.   Using these methods -- which do not come to us effortlessly, but require nurture and vigilance --  we can accomplish wonders.

 \section{Postcript: Reflections}
 
That was the conclusion of the lecture as I gave it.  I'd like to add, in this written version, a few personal reflections.

 \subsection{Thanks}
 
Before concluding I'd like to distribute thanks.  

First I'd like to thank my parents, who cared for my human needs and   encouraged my curiosity from the beginning.  They were children of immigrants from Poland and Italy, and grew up in difficult circumstances during the Great Depression, but managed to emerge as generous souls with an inspiring admiration for science and learning.   I'd like to thank the people of New York, for supporting a public school system that served me extremely well.  I also got a superb undergraduate education, at the University of Chicago.  In this connection I'd especially like to mention the inspiring influence of Peter Freund, whose tremendous enthusiasm and clarity in teaching a course on group theory in physics was a major influence in nudging me from pure mathematics toward physics.   

Next I'd like to thank the people around Princeton who contributed in crucial ways to the circumstances that made my development and major work in the 1970s possible.   On the personal side, this includes especially my wife Betsy Devine.   I don't think it's any coincidence that the beginning of my scientific maturity, and a special surge of energy, happened at the same time as I was falling in love with her.   Also Robert Shrock and Bill Caswell, my fellow graduate students, from whom I learned a lot, and who made our extremely intense life-style seem natural and even fun.    On the scientific side, I must of course thank David Gross above all.  He swept me up in his drive to know and to calculate, and through both his generous guidance and his personal example started and inspired my whole career in physics.   The environment for theoretical physics in Princeton in the 1970s was superb.   There was an atmosphere of passion for understanding, intellectual toughness, and inner confidence whose creation was a great achievement.   Murph Goldberger, Sam Treiman, and Curt Callan especially deserve enormous credit for this.   Also Sidney Coleman, who was visiting Princeton at the time, was very actively interested in our work.  Such interest from a physicist I regarded as uniquely brilliant was inspiring in itself; Sidney also asked many challenging specific questions that helped us come to grips with our results as they developed.   Ken Wilson had visited and lectured a little earlier, and his renormalization group ideas were reverberating in our heads.  

Fundamental understanding of the strong interaction was the outcome of decades of research involving thousands of talented people.   I'd like to thank my fellow physicists more generally.  My theoretical efforts have been inspired by, and of course informed by, the ingenious persistence of my experimental colleagues.  Thanks, and congratulations, to all.   Beyond that generic thanks I'd like to mention specifically a trio of physicists whose work was particularly important in leading to ours, and who have not (yet?) received a Nobel Prize for it.  These are Yoichiro Nambu, Stephen Adler, and James Bjorken.   Those heroes advanced the cause of trying to understand hadronic physics by taking the concepts of quantum field theory seriously, and embodying them in specific mechanistic models, when doing so was difficult and unfashionable.   I'd like to thank Murray Gell-Mann and Gerard  'tHooft for not quite inventing everything, and so leaving us something to do.   And finally I'd like to thank Mother Nature for her extraordinarily good taste, which gave us such a beautiful and powerful theory to discover. 

This work is supported in part by funds provided by the U.S. Department
of Energy (D.O.E.) under cooperative research agreement DE-FC02-94ER40818.

\subsection{A Note to Historians}

I have not, here, given an extensive account of my personal experiences in discovery.  In general, I don't believe that such accounts, composed well after the fact, are reliable as history.   I urge historians of science instead to focus on the contemporary documents; and especially the original papers, which by definition accurately reflect the understanding that the authors  had at the time, as they could best articulate it.   From this literature, it is I think not difficult to identify where the watershed changes in attitude I mentioned earlier occurred, and where the outstanding paradoxes of strong interaction physics and quantum field theory were resolved into modern paradigms for our understanding of Nature.


\begin{thebibliography}{99}
\bibitem{disc} In view of the nature and scope of this write-up, its footnoting will be light.  Our major original papers \cite{gw1, gw2, gw3} are carefully referenced.
\end{thebibliography}
 \end{document}